\documentclass[11pt, a4paper]{revtex4}%
\usepackage{amsfonts}
\usepackage{amsmath}
\usepackage{amssymb}
\usepackage[dvipdfmx]{graphicx}%
\providecommand{\U}[1]{\protect\rule{.1in}{.1in}}

\begin{document}

\title{Toughening in a nacre-like soft-hard layered structure due to weak
nonlinearity in the soft layer}
\author{Yuko Aoyanagi and Ko Okumura\\Department of Physics and Soft Matter Center, Graduate School of Ochanomizu
University, 2--1--1, Otsuka, Bunkyo-ku, Tokyo 112-8610, Japan }

\begin{abstract}
Recently, it has been found experimentally that hydrated nacre exhibits a
nonlinear mechanical response. While mechanical nonlinearity has been shown to
be important in other biological structures, such as spider webs, the
implications of mechanical nonlinearity in nacre have not been explored. Here,
we show that the nonlinear mechanical response of nacre can be reproduced by
an analytical model, which reflects a nacre-like layered structure, consisting
of linear-elastic hard sheets glued together by weakly nonlinear-elastic soft
sheets. We develop scaling analysis on this analytical model, and perform
numerical simulations using a lattice model, which is a discrete counterpart
of the analytical model. Unexpectedly, we find the weak nonlinearity in the
soft component significantly contributes to enhancing toughness by
redistributing the stress at a crack tip over a wider area. Beyond
demonstrating a mechanism that explains the unusual properties of biological
nacre, this study points to a general design principle for constructing tough
composites using weak nonlinearity, and is useful as a guiding principle to
develop artificial layered structures mimicking nacre.

\end{abstract}
\date{\today}
\maketitle

\section{Introduction}

Natural materials often exhibit remarkable hierarchical structures leading to
outstanding mechanical characteristics
\cite{GaoARMR10,Fratzl2007Review,Gao2003PNAS,OkumuraMRS2015,Hellmich2004Bone},
as observed in bone, spider silk \cite{Vollrath2001NatureReview}, and the
exoskeletons of crustaceans \cite{Club2012ScienceA,Lobster2010AM}. Nacre,
which is found in many seashells and protects shellfish from their
environment, is composed of soft and hard layers, and has been studied as a
prototype material for several decades
\cite{Currey1977,JacksonVincentTurner,Sarikaya}. Researchers have been
inspired by nacre's remarkable structure to develop new materials that
demonstrate excellent mechanical performance.
\cite{KatoNacre2000AM,Mayer2005Science,Tomsia2008Science,BioinspiredPolymerFilmScience2008,Science2007LayerPolymer,bouville2014strong,sakhavand2015universal}%
.

There have been many studies discussing the mechanisms responsible for the
toughness of nacre. Toughening mechanisms have been proposed based on a
variety of experimental observations, such as (1) step-wise elongation
\cite{Smith1999NatureNacre}, (2) micro-cracking and crack bridging
\cite{Kamat2000NatureNacre}, (3) thin compressive layers \cite{Rao1999Science}%
, (4) rough layer interfaces \cite{SuoEvansHuchinson}, (5) mineral bridges
\cite{SongBai2003}, and (6) wavy surface of the plates
\cite{Barthelat2007JMPS}. On the theoretical side, various approaches have
been explored, including (1) elastic models \cite{Rao1999Science} based on
analytical solutions \cite{OkumuraPGG2001,Okumura2002nacre} and on scaling
arguments \cite{DeGennesOkumuraCR,okumura2005JPCM}, (2) viscoelastic models
\cite{2003parallel,Okumura2002nacre}, (3) micro-mechanical models
\cite{Kotha2001Micromechanical}, (4) numerical models including finite-element
models \cite{Katti,JiGao2004,Barthelat2007JMPS}, (5) a fuse network model
\cite{Nukala}, and (6) a model with a periodic Young's modulus
\cite{fratzl2007periodic}.

In general, materials start breaking from the tip of small cracks where stress
is concentrated. Therefore, the reduction of the stress concentration at crack
tips is key for material toughness \cite{Griffith,Lawn,Anderson}. Using a
simple linear model of nacre, and deriving an analytical expression of the
stress and strain fields near a crack tip, we have shown that there is a
significant reduction in stress near crack tips in nacre
\cite{OkumuraPGG2001,Hamamoto2008PRE}. This reduction in stress concentration
was numerically confirmed using a simple network model \cite{Aoyanagi2009PRE},
as well as finite-element calculations \cite{Hamamoto2013}. These numerical
studies elucidated a physical picture for the mitigation of stress
concentration, where enhanced elongation of soft layers effectively suppresses
the deformation of the hard layer component, leading to a reduction in stress
concentration. This suppression occurs effectively by the fact that the soft
layers constitute only a small percentage of the bulk material, and thus
stress in the bulk material is governed by the hard layers.

The mechanism of stress reduction above was established using linear models,
but the importance of nonlinear response in biological materials has recently
attracted greater interest. While we previously showed a simple linear model
demonstrates the high mechanical adaptability of spider webs
\cite{AoyanagiOkumura2010}, the nonlinear response of spider silk was shown to
be a key factor in their mechanical superiority \cite{Buehler2012Nature}.
Previously, we have shown the mechanism of stress reduction in nacre using a
linear model. Recently, it been shown experimentally that nacre also exhibits
a nonlinear mechanical response \cite{NanoLett2004Nacre, Barthelat2007JMPS,
Kakisawa2008Nacre}, as spider silk does.

Here we construct a nonlinear model of a nacre-like structure that generalizes
the linear model studied in Ref. \cite{OkumuraPGG2001} and reflects the recent
experimental results in Ref. \cite{NanoLett2004Nacre,
Barthelat2007JMPS,Kakisawa2008Nacre}. Using this model, we perform simulations
and derive scaling laws for a sample with a line crack. The numerical
simulations, and scaling laws derived from this model show that the stress
concentration near the crack tips is reduced significantly in the nonlinear
case when the stress concentration is significant. The scaling laws
demonstrate that toughness and strength are enhanced by a common factor that
elucidates simple design principles for developing artificial materials
mimicking nacre.

\section{Results}

\subsection{Scaling analysis of our model for nacre}

\begin{figure}[h]
\includegraphics[width=1.\linewidth]{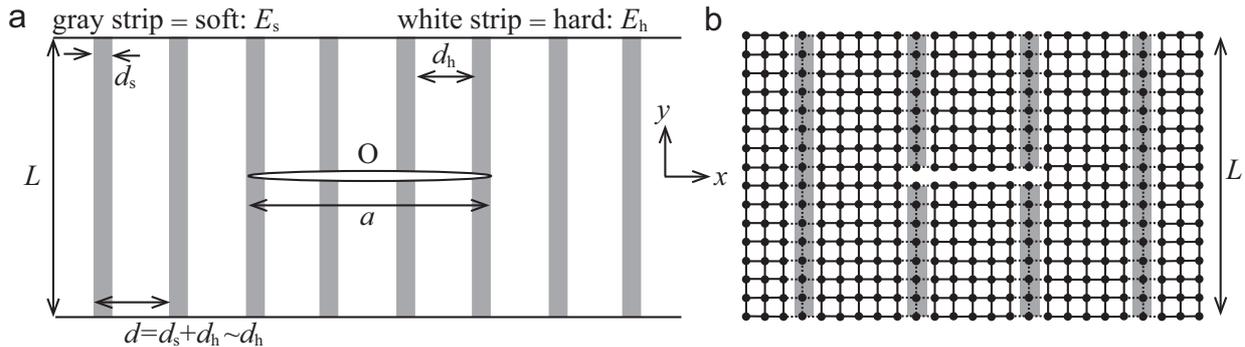}\caption{(a) Layered structure
with a crack of a finite length. Gray stripes correspond to soft thin layers.
(b) Lattice model for numerical simulation with a crack in the middle to be
stretched in the $y$ direction, where the crack tips are located at soft
layers ($d_{s}=d_{0},d=8d_{0}$, $a=8d_{0}$ with $d_{0}$ the mesh size in the
illustration).}%
\label{fig1}%
\end{figure}

\label{s1-1} In our analytical model, hard and thick layers of thickness
$d_{h}$ are glued together with soft and thin layers of thickness $d_{s}$
($d_{h}>>d_{s}$), with a period
\begin{equation}
d=d_{s}+d_{h}\simeq d_{h}, \label{e1}%
\end{equation}
as illustrated in Fig.~\ref{fig1}(a) (The corresponding lattice model for
numerical simulation is illustrated in (b), which will be explained below in
Sec. \ref{D1}). In our nonlinear model, the relation between the
characteristic magnitude of stress and strain ($\sigma$ and $e$) are given by
\begin{align}
\sigma &  \simeq E_{h}e^{1/n_{h}}\text{ \ (for hard layers)}\nonumber\\
\sigma &  \simeq E_{s}e^{1/n_{s}}\text{ \ (for soft layers).} \label{e2}%
\end{align}
Here, the nonlinear exponent ($n_{s}$ or $n_{h}$) is positive, and the elastic
modulus is denoted $E_{h}$ for the hard layers and $E_{s}$ for the soft layers.

As seen below, it is convenient to introduce a small parameter $\varepsilon$
defined as%
\begin{equation}
\varepsilon=\frac{E_{s}}{E_{h}}\left(  \frac{d_{h}}{d_{s}}\right)  ^{1/n_{s}}.
\label{e03}%
\end{equation}
In our analytical model, we consider the small $\varepsilon$ limit and we are
not interested in the exact values but just the orders of magnitude of the
physical quantities.

In the small $\varepsilon$ limit, we show below that the dominant components
of the stress and strain fields satisfy the following relations:
\begin{equation}
\sigma_{yy}\sim E_{h}e_{yy}^{1/n_{h}}\text{ \ \ and \ \ }\sigma_{yx}%
\sim\varepsilon E_{h}e_{yx}^{1/n_{s}} \label{e09}%
\end{equation}
Here, note that the Young's and shear moduli (e.g., of hard layers) are of the
same order of magnitude ($\sim E_{h}$) because they only differ by a numerical
coefficient on the order of one. The continuum strain field is defined as
$e_{ij}=(\partial_{i}u_{j}+\partial_{j}u_{i})/2$ for $(x_{1,}x_{2}%
,x_{3})=(x,y,z)$. Note that such continuum fields are valid in the present
layered structure only on scales larger than the period $d$.
Equation~(\ref{e09}) is a natural nonlinear extension of the linear relation
obtained for the simple model of nacre discussed in Ref. \cite{OkumuraPGG2001}.

Now we derive Eq.~(\ref{e09}) by considering the elemental modes of
deformation appropriate for shear ($\sigma_{yx}$) and stretch ($\sigma_{yy}$)
in the $y$ direction. When the composite is sheared in the $y$ direction, only
the soft layers are stretched, and $e_{yx}\sim(d_{s}/d)e_{yx}^{(s)}$. From
Eq.~(\ref{e2}), this results in $\sigma_{yx}\sim\varepsilon E_{h}%
e_{yx}^{1/n_{s}}$. In contrast, when the composite is stretched in the $y$
direction, both the soft and hard layers are stretched so the stress is
dominated by the response from the hard layers, and $\sigma_{yy}\sim
E_{h}e_{yy}^{1/n_{h}}$. Here, we have not considered any slip or separation
between the interfaces of the layers, because we are interested in deriving
the critical condition for failure. The mutual sliding of hexagonal hard
platelets observed experimentally in a study on the toughness of nacre
\cite{NanoLett2004Nacre} is regarded as a state that exists after the critical
condition for failure is satisfied and is beyond the scope of our description.

We consider a composite governed by Eq. (\ref{e09}) with a crack running in
the $x$ direction. The composite is stretched in the $y$ direction as shown in
Fig.~\ref{fig1}(a), where the plane strain or plane stress condition is
satisfied (the sample is thick or thin in the $z$ direction). Thus, the
dominant stress components are those pointing in the $y$ direction, or
$\sigma_{yy}$ and $\sigma_{yx}$. This is the reason we have considered only
these components in the above.

As explained in the Appendix, we can derive the following scaling laws for the
fracture toughness and strength of the composite in the presence of a large
crack when $\varepsilon$ is small:%

\begin{align}
G  &  =\lambda G_{h}\label{em1}\\
\sigma_{F}  &  =\lambda^{\frac{1}{1+n_{h}}}\sigma_{F}^{MN} \label{em2}%
\end{align}
where the common enhancement factor $\lambda$ is given by%
\begin{equation}
\lambda\sim\frac{d}{a_{0}}\varepsilon^{-\frac{1}{1+1/n_{s}}}\left(
\frac{G_{h}/E_{h}}{a_{0}}\right)  ^{\frac{1}{1+n_{h}}+\frac{1}{1+1/n_{s}}-1}
\label{em3}%
\end{equation}
Here, $G_{h}$ and $\sigma_{F}^{MN}$ are the fracture toughness (fracture
surface energy) and strength of a brittle monolithic hard material,
respectively, and $a_{0}$ is the size of the so-called Griffith cavity, as
discussed below. These scaling laws can be viewed as natural nonlinear
extensions of results obtained from an analytical solution in Ref.
\cite{OkumuraPGG2001} (see Appendix).

As explained in the Appendix, from Eqs. (\ref{em1}) and (\ref{em2}), we can
derive the following expression for the maximum stress that appears near a tip
of a large crack in a network model of system size $L$ when the remote stress
$\sigma_{0}$ is applied at the top and bottom edges of the sample:%

\begin{equation}
\frac{\sigma_{M}}{E_{h}}\sim\left(  \varepsilon\left(  \frac{\sigma_{0}}%
{E_{h}}\right)  ^{\left(  1+n_{h}\right)  \left(  1+1/n_{s}\right)  }\left(
\frac{L}{d}\right)  ^{1+1/n_{s}}\right)  ^{\frac{1}{\left(  1+n_{h}\right)
+\left(  1+1/n_{s}\right)  }} \label{em5}%
\end{equation}
The physical implications of Eqs. (\ref{em3}) and (\ref{em5}) will be
discussed later.

\subsection{Relevance of our analytical model to previous experiments}

\label{s1-2} 
\begin{figure}[h]
\includegraphics[width=0.5\linewidth]{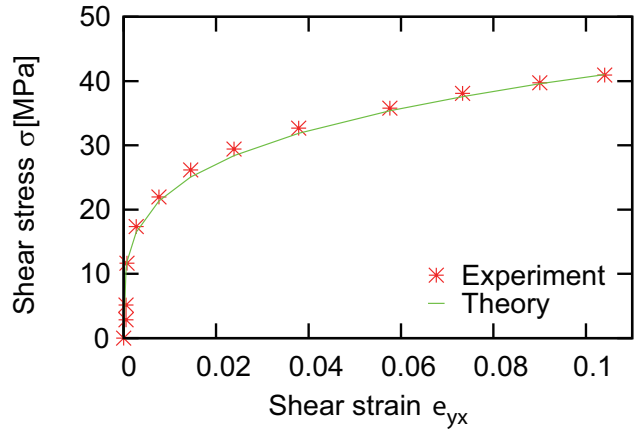}\caption{Comparison of an
experimentally obtained stress-strain curve (red star symbols) for shear
deformation in \cite{Barthelat2007JMPS} with a stress-strain curve obtained
with our analytical model (solid line) shows close agreement.}%
\label{Fig2}%
\end{figure}

In this section, we show the relevance of our analytical model
developed above to biological nacre by showing that Eq. (\ref{e09}) with
appropriate parameters well reproduces previous experiments. Experimental
studies have shown the tensile elastic response of the hard layers in nacre is
linear with a modulus, typically around 65 GPa (before microscopic internal
failure), as confirmed by nano-indentation \cite{NanoLett2004Nacre} and
tension tests \cite{Barthelat2007JMPS}. This means that Eq. (\ref{e09}) with
$n_{h}=1$ and $E_{h}\simeq65$ GPa reproduces experiments well. Additionally,
the shear response of the wet composite has recently been obtained
\cite{Barthelat2007JMPS}, and an experimental stress-strain curve is shown by
the star symbols in Fig. \ref{Fig2}. This curve can be reproduced at a
semi-quantitative level by Eq. (\ref{e09}) with $\varepsilon E_{h}\simeq72$
MPa and $n_{s}=4$, as shown by the solid curve in Fig. \ref{Fig2}. This means
that, in biological nacre, the soft layer is weakly nonlinear with
$n_{s}\simeq4$, and $\varepsilon$ is significantly small with $\varepsilon$
$\simeq0.001$ ($\varepsilon E_{h}\simeq72$ MPa and $E_{h}\sim65$ GPa result in
$\varepsilon\sim0.001$).

The estimates, $E_{h}\sim65$ GPa, $\varepsilon E_{h}\simeq72$ MPa, and
$n_{s}\simeq4$, imply that the soft layer is very soft with $E_{s}\simeq35$
MPa. This is because the overall volume of the soft layer is typically 5 per
cent, implying $d_{s}/(d_{s}+d_{h})\simeq1/20$ with $d_{h}$ typically 0.5
$\mu$m, and the parameter set $E_{s}\simeq35$ MPa, $E_{h}\sim65$ GPa,
$d_{s}/d_{h}=1/19$, and $n_{s}=4$ in Eq. (\ref{e03}) gives $\varepsilon
E_{h}\sim72$ MPa. Note that, although the measurement of the force response of
the soft component is difficult because of extremely small sample sizes of the
soft component, it has been known that the soft component behaves like a soft
gel \cite{Smith1999NatureNacre,Kakisawa2008Nacre,Meyers2008NacreSoft}.

\subsection{Physical implications of the scaling laws}

\begin{figure}[h]
\includegraphics[width=1.\hsize]{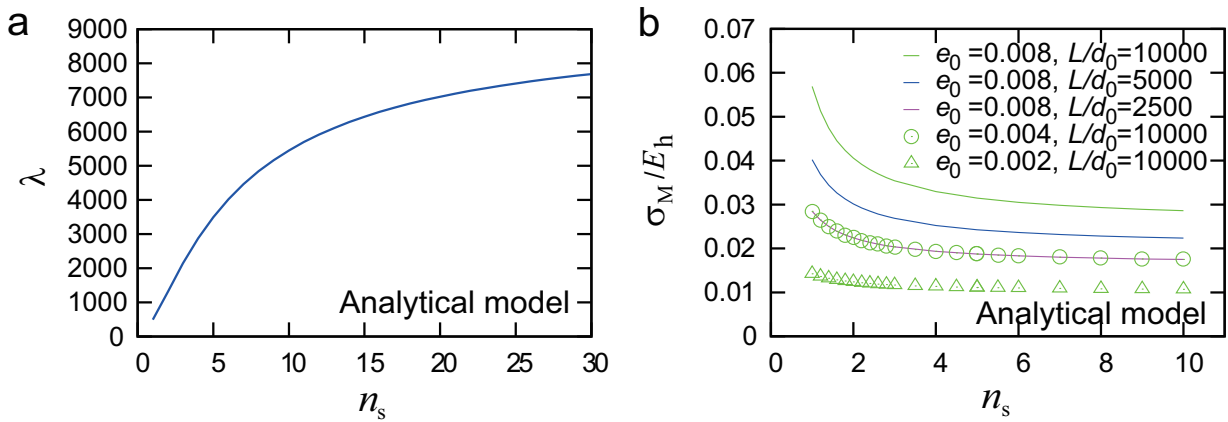} \caption{(a) The common enhancement
factor for the fracture toughness and strength as a function of $n_{s}$,
obtained from the scaling law in Eq. (\ref{em3}). (b) The maximum stress that
appears at a crack tip as a function of $n_{s}$, obtained from the result of
our scaling analysis given in Eq.(\ref{em5}) setting the numerical factor for
the relation to be 1.}%
\label{Fig3}%
\end{figure}

Here, we discuss the rough orders of magnitude of important parameters in Eq.
(\ref{em3}), to better understand its physical implications. The
characteristic size of cavities in the hard layers, or the Griffith cavity
$a_{0}$, is about $50$ nm, comparable to the thickness of the soft layers
$d_{s}$, or $a_{0}\cong d_{s}$. The cavities or defects in the hard layers may
be created from protein molecules in the soft layers intruding into, and being
trapped within, the mineral crystals during biomineralization process in nacre
\cite{Gao2003PNAS}. The fracture toughness $G_{h}$ is on the order of 10
J/m$^{2}$, which leads to $G_{h}/E_{h}\sim0.1$ nm. In summary, typical values
of the important factors in the common enhancement factor $\lambda$ can be
given as
\begin{equation}
d/a_{0}=50,\frac{G_{h}/E_{h}}{a_{0}}=1/100,\frac{E_{s}}{E_{h}}=\frac
{35}{65000},\frac{d_{s}}{d_{h}}=1/19. \label{eqOrd}%
\end{equation}
Note that this parameter set means $\varepsilon E_{h}\sim72$ MPa at $n_{s}=4$.

In Fig. \ref{Fig3}(a), we show Eq. (\ref{em3}) with the parameter set in Eq.
(\ref{eqOrd}). The enhancement factor $\lambda$ as a function of $n_{s}$
sharply rises up until $n_{s}\sim5-10$ and then tends to plateau. This implies
that a relatively weak nonlinearity is sufficient for toughening. Note that in
Fig. \ref{Fig3}(a), the order of the toughness enhancement $\lambda$ is
predicted in the thousands, which is consistent with previous experiments
\cite{JacksonVincentTurner}.

The same physical conclusion can be obtained from Eq. (\ref{em5}) when the
ratio $L/d$ and the remote strain $e_{0}$ are relatively large, where
$e_{0}=\sigma_{0}/E_{h}$ in the present case of $n_{h}=1$. In Fig.
\ref{Fig3}(b), Eq. (\ref{em5}) is plotted at $E_{s}/E_{h}=35/65000$ and
$d_{s}/d_{h}=1/19$ for different values of $L/d$ and $e_{0}$. Equation
(\ref{em5}) represents the maximum stress at the crack tip, and is a measure
of strength because a material is considered to be strong when the value of
this quantity is small. As seen in Fig. \ref{Fig3}(b), this quantity drops
sharply up until $n_{s}\sim5$ and then it approaches a plateau when $e_{0}$
and $L/d$ are relatively large, which suggests that weak nonlinearity is
sufficient for toughening. (The sharp drop in maximum stress is less
pronounced when $e_{0}$ and $L/d$ are smaller, and this behavior is physically
interpreted in the Discussion in terms of the degree of stress concentration.)

\subsection{Numerical simulation}

\subsubsection{Lattice model}

\label{D1} We performed numerical simulation using a nonlinear model extended
from a linear two-dimensional network model of nacre studied in Ref.
\cite{Aoyanagi2009PRE}. The lattice model for this simulation is illustrated
in Fig.~\ref{fig1}(b). We consider a grid system in which nodal points are
connected by springs that reflect the nonlinearity specified in Eq.~(\ref{e2}%
). The values of $E_{s}$ and $E_{h}$ in our simulation are set as $E_{h}=65$
GPa and $E_{h}=35$ MPa, with $d_{h}/d_{s}=19$, as in Eq. (\ref{eqOrd}), and
with $n_{h}=1$ and $n_{s}=4$ to mimic nacre. (See Appendix for further details).

To study the strength of the system against failure, we introduce a line crack
in the $x$ direction in a stretched system, and quantify how the stress is
concentrated around the crack tips for a fixed deformation at the top and
bottom edges of the sample at mechanical equilibrium. The two crack tips are
placed at soft layers. This is because the critical condition for failure on a
scale larger than the layer period $d$ should be the condition whether a crack
develops further when the crack tip is stopped at a soft layer. We calculate
the positions of the beads when the elastic energy is minimized by the
conjugate gradient method \cite{NumericalRecipes}.

For a systematic comparison, we consider three models, (1) a nonlinear layered
model which mimics natural nacre with $\varepsilon\ll1$ and $(n_{h}%
,n_{s})=(1,4)$, (2) a linear layered model, which is different from the
natural nonlinear model only in that $n_{s}$ is set to one, and (3) a
monolithic model ($\varepsilon=1$), which is different from the linear model
only in that the value of $E_{s}$ is raised to match that of $E_{h}$. We set
the remote strain $e_{0}$ to be $0.008$ in each simulation, which is one of
the values of $e_{0}$ used in Fig. \ref{Fig3}(b).

\subsubsection{Stress distribution in the presence of a crack}

\begin{figure}[h]
\includegraphics[clip, width=0.8\linewidth]{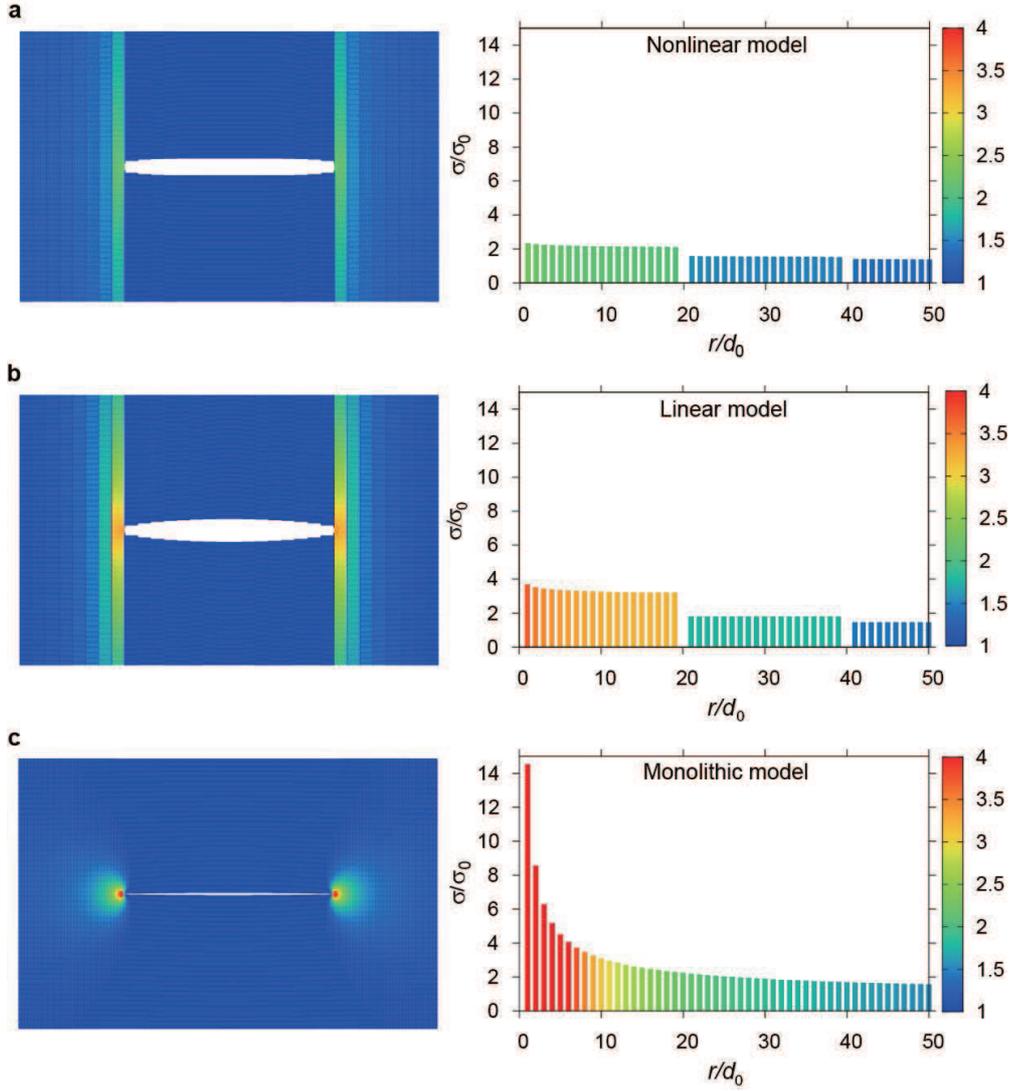}\caption{Comparison of
stress distribution with the three models: (a) Nonlinear model mimicking
nacre. (b) Linear model. (c) Monolithic model. (left) Stress distribution
around a crack in the network model composed of $960\times10000$ grid points.
(right) Stress along the $x$ axis as a function of distance from the crack
tip.}%
\label{Fig4}%
\end{figure}

In Fig.~\ref{Fig4}(left), a comparison of the stress distribution around a
line crack is given for the three models with a crack of the same size, where
the maximum of the color scale (red) is set to the maximum stress that appears
at the crack tips in the linear model. By comparing the three distribution
maps on the left, we can clearly see the stress concentration near the crack
tips is significantly reduced and delocalized in the layered cases (a and b),
and reduced and delocalized most in the nonlinear case in (a).

To quantify the stress near the crack tips, we plot how the stress changes
around one of the crack tips along the $x$ axis in Fig.~\ref{Fig4}(right) to
more clearly see how the stress concentration is minimized in the nonlinear
case. As expected, in all the three models, the maximum stress appears at the
crack tip, or more precisely, at $r/d_{0}=1$, where $r$ is the distance from
the crack tip and $d_{0}$ is the mesh size of the network model. In the
nonlinear model, the enhancement factor for the crack tip stress compared with
the remote value $\sigma_{0}$ is rather small, and is approximately $2$. The
enhancement factors for the linear model and the monolithic model are
approximately $4$ and $14$, respectively.

The enhancement of the stress at the crack tip significantly increases as we
move from nonlinear to linear to monolithic models. This is in accord with the
physical implications demonstrated in Fig. \ref{Fig3}, that weak nonlinearity
in the soft layer enhances the strength of the layered structure.

\subsection{Consistency of our scaling analysis and numerical simulation}

In this section, we demonstrate that the results of the numerical simulation
semi-quantitatively agree with the scaling analysis. Note that the lattice
model for numerical simulation should reduce to the analytical model in the
continuum limit.

Equation (\ref{em5}) predicts the correct order of the maximum stress at the
crack tip in the three numerical models in Fig. \ref{Fig4}(right). The ratios
of the maximum stress for nonlinear, linear composite and monolithic models
are calculated as $1:2:6$, respectively, from Eq. (\ref{em5}), while the
corresponding ratios are given as $1:2:7$ from Fig. \ref{Fig4}(right). This
level of agreement is remarkable, considering that the assumptions made in the
derivation of the scaling laws are only reasonably well satisfied in our
simulation for numerical limitations (see Appendix for the details). In this
way, we conclude that Eq. (\ref{em5}), and thus Eqs. (\ref{em1}) and
(\ref{em2}), are consistent with and supported by the results of our numerical
calculation in Fig. \ref{Fig4}.

\begin{figure}[h]
\includegraphics[width=0.5\hsize]{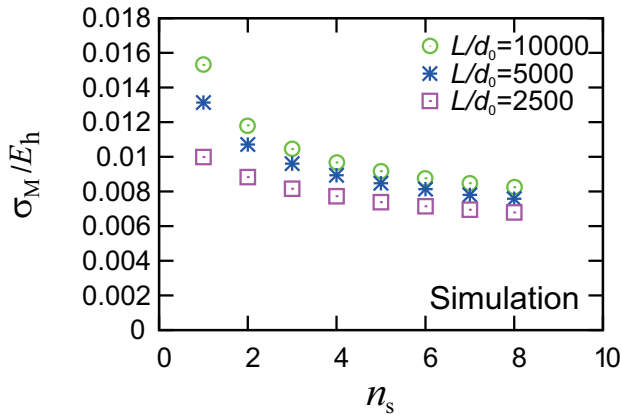} \caption{The maximum stress that
appears at a crack tip as a function of the nonlinear index $n_{s}$ for
different system size $L$ obtained from numerical calculation.}%
\label{Fig5}%
\end{figure}

To further confirm semi-quantitative agreement between our simulation and
scaling analysis, we plot the maximum stress that appears at a crack tip as a
function of the nonlinear index $n_{s}$ in Fig. \ref{Fig5}. The three curves
in Fig. \ref{Fig5} may be compared with the three solid curves in Fig.
\ref{Fig3}(b), because they were obtained at the same strain $e_{0}=0.008$ and
for the same three values of $L/d$. As expected, the curves in each figure
semi-qualitatively agree, further supporting the scaling analysis.

\section{Discussion}

A crucial factor for toughening is stress concentration, and the degree of
stress concentration is governed clearly by the remote stress $e_{0}%
=\sigma_{0}/E_{h}$ and less clearly by the quantity $L/d$. An increase in
$L/d$ results in an increase in stress concentration. This is because the
stress at a distance $r$ from the crack tip is given as $\sigma(r)\simeq
\sigma_{0}(L/r)^{1/(1+n)}$ for a non-linear system governed by the
stress-strain relation $\sigma\simeq Ee^{1/n}$, and the maximum stress that
appears at the crack tip is given by a cutoff value $\sigma_{0}(L/d)^{1/(1+n)}%
$ for a system with a scale $d$ below which the continuum description breaks
down \cite{aoyanagi2017}.

The fact that both $\sigma_{0}/E_{h}$ and $L/d$ control the stress
concentration is reflected in Eq. (\ref{em5}). Analytically, the functional
dependences of Eq. (\ref{em5}) on $e_{0}$ and $L/d$ are very similar. In fact,
as demonstrated numerically in Fig. \ref{Fig3}(b), the initial drop in the
maximum stress as a function of $n_{s}$ becomes significant as $L/d$
increases, and the same is true as $e_{0}$ increases.

This change with $e_{0}$ of the initial drop in the maximum stress with
$n_{s}$ is physically understood from the fact that, if we simply plot
$\sigma=Ee$ and $\sigma=Ee^{1/n}$ with $n$ larger than unity, the nonlinear
stress ($Ee^{1/n}$) is smaller than the linear stress ($Ee$) when the strain
$e$ is larger than unity (the nonlinear stress is larger when $e$ is smaller
than unity). This suggests that nonlinearity in the soft layer contributes to
stress reduction effectively when strain acting on soft layers near the crack
tip is large. This happens even if the remote strain acting on the composite
is considerably small. Note that the strain in soft layers is significantly
larger than the strain in the soft-hard composite near a crack tip where shear
deformation is dominant. Even if the remote strain acting on the composite is
small, shear strain of soft layers near a crack tip can be large, as suggested
in the paragraph below Eq. (\ref{e09}). (This significantly enhanced
deformation in soft layers near a crack tip is demonstrated in previous
numerical calculations \cite{Aoyanagi2009PRE,Hamamoto2013}.) In summary, when
the stress concentration is high, its reduction due to nonlinearity becomes greater.

The initial drop in the maximum stress as a function of $n_{s}$ is one
advantage of weak nonlinearity, and this advantage is more effective as the
stress concentration becomes higher, or as $e_{0}$ and/or $L/d$ increases.
This behavior of the maximum stress was demonstrated in Fig. \ref{Fig3}(b),
and can physically understood as explained above.

Another advantage of weak nonlinearity is the increase in the enhancement of
$\lambda$ with an increase in $n_{s}$. This is predicted from Eq. (\ref{em3}),
as demonstrated in Fig. \ref{Fig3}(a), and this expression is obtained in the
limit of large $L/d$, or in the case where the stress concentration is
significant. This implies that if the stress concentration at a crack tip is
low, this second advantage of weak nonlinearity predicted in Eq. (\ref{em3})
may be diminished. This is expected because stress-concentration reduction due
to nonlinearity in soft layers (the first advantage) is more efficient when
the stress concentration at a crack tip is high, and stress-concentration
reduction and the enhancement factor $\lambda$ should be positively
correlated. In summary, the two advantages of weak nonlinearity, originating
in Eqs. (\ref{em3}) and (\ref{em5}), are physically the same, and the strength
of the advantages enhance as $e_{0}$ and $L/d$ increase, or stress
concentration increases.

In the case of biological nacre, it has been reported that hard sheets tend to
be separated into hexagonal tablets at a significantly low strain of around
0.001 \cite{Barthelat2007JMPS}. If this is always the case, then the two
advantages of weak nonlinearity may not be so significant in real, biological
nacre. This is because as seen in Fig. \ref{Fig3}(b) the first advantage of
weak nonlinearity is visible at $e_{0}=0.008$, and the two advantages tend to
decrease as $e_{0}$ decreases (the curve for $e_{0}=0.002$ in Fig.
\ref{Fig3}(b) shows only a slight initial drop). Note that toughness and
strength are governed by the strength of the two advantages at the critical
state of failure when $e_{0}$ reaches the threshold strain for separation of tablets.

However, the value of around 0.001 reported in \cite{Barthelat2007JMPS} is
just one sample of nacre, and other natural specimens may vary. If other
samples of nacre possess threshold strains which are fairly larger than 0.001,
and their mechanical responses are explained by weak nonlinearity in the soft
layers, then the weak nonlinearity could be interpreted as a mechanical
optimization attained by biological nacre. This is because, if the threshold
strain is fairly large, at the critical state of failure, the initial drop in
the maximum stress at the crack tip as a function of $n_{s}$ is rather steep,
and the maximum stress approaches to a plateau at a small $n_{s}$, and thus
weak nonlinearity is sufficient to obtain a high strength of the two advantages.

At any rate, since the present study demonstrates that the advantages of weak
nonlinearity in the soft layer is pronounced for a reasonably small strain of
0.008, this mechanism could be useful for developing artificially tough
layered structures mimicking nacre, where hard plates do not break into
tablets. Note that reducing the strain below 0.008 in numerical simulation is
technically difficult due to precision in calculations.

\section{Conclusion}

We demonstrated that in natural nacre the hard layers are linear and the soft
layers are weakly nonlinear ($n_{s}\sim4$) on the basis of recent experimental
results, and studied mechanical properties on the basis of nonlinear models
reflecting this nonlinearity both analytically and numerically. The
combination of numerical simulation and scaling arguments led to the following
conclusions for a reasonably small strain: (1) a nonlinear model significantly
contributes to reduce the stress concentration at a crack tip, compared with a
linear model, and (2) a weak nonlinearity, as selected in biological nacre, is
sufficient for this reduction.

We analytically derived the enhancement factor $\lambda$ common to both the
fracture strength and toughness. This factor, given in Eq. (\ref{em3}), shows
that the mechanical superiority of nacre-like layered structures is controlled
by the factors $d/a_{0}\sim d_{h}/d_{s}$ and $\varepsilon^{-1}=(E_{h}%
/E_{s})(d_{s}/d_{h})^{1/n_{s}}$, which are larger than one. These factors
elucidate two design principles for the enhancement: (1) the soft-hard
combination ($E_{s}<E_{h}$) and (2) the thin-thick combination with weak soft
nonlinearity ($d_{s}<d_{h}$ and $n_{s}\gtrsim1$). These principles imply that
the soft-hard combination, hierarchical structure, and nonlinearity are all
important for developing tough structures mimicking nacre.

\section*{Acknowledgement}

This research was partly supported by Grant-in-Aid for Scientific Research (A)
(No. 24244066) of JSPS, Japan, and by ImPACT Program of Council for Science,
Technology and Innovation (Cabinet Office, Government of Japan). We thank
Professor Edward Foley (Ochanomizu University) for editing the manuscript.

\section*{Appendix}

\subsection{Simulation}

\label{A1}

The simulation model is composed of $960\times10000$ grid points ($L=10000$ in
the unit $d_{0}$), and the system size in the non-stretched state is
960$d_{0}\times10000d_{0}$, with $d_{s}=d_{0}$ and $d=20d_{0}$ ($d_{h}%
=19d_{0}$). We introduced a line crack of size $a=320d_{0}$ (whose width is
$d_{0}$) with the tips located at soft layers. We stretched the system in the
$y$ direction perpendicular to the line crack. The strain in our simulation is
set to 0.008, which corresponds to the experimental value of tensile strain on
nacre at failure \cite{Barthelat2007JMPS}, with $E_{h}/E_{s}=65000/35$ in the
nonlinear and linear layered models, and $E_{h}/E_{s}=1$ in the monolithic
model. Note that assumptions in the scaling theory, e.g., $L,a\gg d\gg d_{0}$,
is relaxed (or only marginally satisfied) in the simulation due to practical
limitations on numerical calculation.

\subsection{Analytical model}

The scaling structure of the elastic energy (per volume) can be given as
\begin{equation}
F\sim E_{h}e_{yy}^{1/n_{h}+1}+\varepsilon E_{h}e_{yx}^{1/n_{s}+1} \label{e02}%
\end{equation}
where the numerical coefficients of the both terms are set to one for
simplicity. Equation~(\ref{e02}) is obtained from the expression
$F=(\sigma_{yy}e_{yy}+\sigma_{yx}e_{yx})/(n+1)$, together with Eq.~(\ref{e09}).

Equation~(\ref{e02}) is a natural nonlinear extension of the energy, first
derived in Ref. \cite{OkumuraPGG2001}. On the basis of this expression, we
develop scaling arguments to derive Eqs. (\ref{em1}), (\ref{em2}), and
(\ref{em5}) in the text. These expressions can be viewed as a natural
nonlinear extension of those obtained in Ref. \cite{OkumuraPGG2001}, and by
setting $n_{s}=n_{h}=1$ in Eqs. (\ref{em1}), (\ref{em2}), and (\ref{em5}), we
can reproduce the corresponding expressions obtained in Ref.
\cite{OkumuraPGG2001} from an exact analytical solution in appropriate limits.
The details of the derivation of the scaling laws are given below.

\subsubsection{Details of Derivation}

To obtain the scaling laws for the nonlinear model of nacre, we consider a
crack problem in a generic way. A crack of size $a$ is located along the $x$
axis in a plate whose size in the $y$ direction is characterized by $L$ (the
sample size in the $x$ direction is comparable to $L$). This plate is in
tension. The following derivation is inspired by the arguments in Ref.
\cite{okumura2005fracture}.

We consider two characteristic length scales, $X$ and $Y$, in the $x$ and $y$
directions. A scaling relation between $X$ and $Y$ is given by balancing the
two terms in the energy in Eq.~(\ref{e02}):%
\begin{equation}
(u/Y)^{1/n_{h}+1}\sim\varepsilon(u/X)^{1/n_{s}+1} \label{e01}%
\end{equation}

The critical condition for fracture is that the elastic energy that is
released as a result of the creation of a crack of size $a$ matches the
fracture energy to create the crack:%
\begin{equation}
E_{h}(u/Y)^{1/n_{h}+1}aY\sim aG \label{e04}%
\end{equation}
Here, the fracture energy per area is denoted $G$. Note that if Eq.~(\ref{e01}%
) holds, the two terms in Eq.~(\ref{e02}) are of the same order.

From Eqs.~(\ref{e01}) and (\ref{e04}), we can express $u$ and $Y$ in terms of
$X$, and from these expressions, we obtain the following expression for the
$(y,y)$ component of the stress at the critical state of failure that scales
as $\sigma\sim E_{h}(u/Y)^{1/n_{h}}\sim G/u$:%
\begin{equation}
\frac{\sigma_{F}}{E_{h}}\sim\left(  \varepsilon\left(  \frac{G/E_{h}}%
{X}\right)  ^{1+1/n_{s}}\right)  ^{\frac{1}{2+n_{h}+1/n_{s}}}\sim\left(
\frac{G/E_{h}}{Y}\right)  ^{\frac{1}{^{1+n_{h}}}} \label{e05}%
\end{equation}

This is a generalized version of a nonlinear Griffith's failure formula that
was proposed theoretically in Ref. \cite{Aoyanagi2009JPSJ} and confirmed
experimentally in Ref.\cite{SoneMoriJSPS}. If we set $\varepsilon=1$ and
$n_{s}=n_{h}$ in Eq. (\ref{e05}) we obtain a nonlinear Griffith's formula for
a monolithic hard material:%
\begin{equation}
\sigma_{F}^{MN}\sim E_{h}\left(  \frac{G_{h}/E_{h}}{X_{h}}\right)  ^{\frac
{1}{1+n_{h}}}. \label{ea05}%
\end{equation}
Here, we have set $G=G_{h}$ and $X=X_{h}$ (in fact, $X_{h}=Y_{h}$ because the
system is isotropic), where $G_{h}$ and $X_{h}$ ($Y_{h}$) stand for the
fracture energy and characteristic scale in the $x$ ($y$) direction for the
monolithic hard material.

The nonlinear Griffith formula in Eq. (\ref{ea05}) is a nonlinear version of
the classic Griffith formula, and in the linear case ($n_{h}=1$), Eq.
(\ref{ea05}) reduces to the well-known expressions for failure stress:
$\sigma\sim\sqrt{E_{h}G/a}$ for $a<L$ and $\sigma\sim\sqrt{E_{h}G/L}$ for
$a>L$.

If we set $X_{h}=a_{0}$ in Eq. (\ref{ea05}), we obtain the intrinsic failure
stress of hard layers when there are no macroscopic cracks:%
\begin{equation}
\sigma_{h}\sim E_{h}\left(  \frac{G_{h}/E_{h}}{a_{0}}\right)  ^{\frac
{1}{1+n_{h}}} \label{e06}%
\end{equation}
Here, $a_{0}$ is the typical size of defects in the layers that play the role
of Griffith cavities.

The scaling structure for the singularity of the stress field at a crack tip
should be given in the form,%
\begin{equation}
\sigma(r)\sim\sigma_{0}(X/r)^{\eta} \label{ea2}%
\end{equation}
where $\sigma_{0}$ and $r$ are the characteristic size of the remote stress
and the distance from the crack tip, respectively. The exponent $\eta$ can be
determined by a principle that, at the critical state of failure [$\sigma_{0}$
is equal to $\sigma_{F}$ in Eq.~(\ref{e05})], the field $\sigma(r)$ becomes
independent of $X$ as we approach the singularity at $r=0$ ($r<<X$)
\cite{Okumura2004DNG}. From this principle, we obtain the desired exponent,
\begin{equation}
\eta=\frac{1+1/n_{s}}{2+n_{h}+1/n_{s}},
\end{equation}
and the singular field at the critical state of failure,
\begin{equation}
\sigma(r)\sim E_{h}\left(  \varepsilon\left(  \frac{G/E_{h}}{r}\right)
^{1+1/n_{s}}\right)  ^{\frac{1}{2+n_{h}+1/n_{s}}}\text{.} \label{e08}%
\end{equation}
Note that this principle allows us to reproduce the well-known crack-tip
singularity obtained by the Hutchinson, Rice and Rosengren (HRR)
\cite{Hutchinson1968,RiceRosengren1968}, and if we set $n_{s}=n_{h}=n$,
$E_{s}=E_{h}$, and $d_{s}=d_{h}$, Eq.~(\ref{ea2}) reduces to the well-known
expression of the HRR singularity: $\sigma(r)\sim\sigma_{0}(X/r)^{\frac
{1}{1+n}}$. This further reduces to the well-known singularity in the linear
case ($n=1$): $\sigma(r)\sim\sigma_{0}\sqrt{X/r}$.

The maximum force allowed at the crack tip is given by Eq.~(\ref{ea2}) at
$r\sim d$. This is expected because the continuum expression should be cut-off
at this scale, and the continuum theory is valid only beyond length scales
larger than the layer period $d$. From this second principle, the maximum
stress at the crack tip at the critical state of failure is given by
Eq.~(\ref{e08}) with $r=d$.

In fact, this second principle is numerically confirmed in Refs.
\cite{Nakagawa,Aoyanagi2009JPSJ,Takahashi2014}. The authors considered a line
crack of size $a$ in a two dimensional nonlinear network system whose mesh
size is $d$, and showed that, as expected, the maximum stress always appears
at the crack tips. Moreover, they showed that the maximum stress follows the
scaling law $\sigma\simeq\sigma_{0}\left(  a/d\right)  ^{1/(1+n)}$ in a clear
way for the remote stress $\sigma_{0}$, whereas the well-known singularity
$\sigma\simeq\sigma_{0}\left(  a/r\right)  ^{1/(1+n)}$ near a crack tip at the
distance $r$ is predicted in Refs. \cite{Hutchinson1968,RiceRosengren1968}.

The critical condition for failure is satisfied when this maximum stress
matches $\sigma_{h}$ in Eq.~(\ref{e06}), and at the critical state, the tip
stress reaches the intrinsic failure stress of hard layers. This matching
condition leads to an expression of the fracture energy $G$:
\begin{equation}
\frac{G}{G_{h}}=\frac{d}{a_{0}}\varepsilon^{-\frac{1}{1+1/n_{s}}}\left(
\frac{G_{h}/E_{h}}{a_{0}}\right)  ^{\frac{1}{1+n_{h}}+\frac{1}{1+1/n_{s}}-1}
\label{e07}%
\end{equation}
The enhancement factor for the fracture energy is given in the left-hand side.

Another scaling relation between $X$ and $Y$ is obtained by considering the
characteristic size of the remote stress $\sigma_{0}$ and the remote strain
$e_{0}$. They satisfy the relationships%
\begin{equation}
\sigma_{0}\sim E_{h}e_{0}^{1/n_{h}}\text{ \ and \ }e_{0}\sim u/Y\text{.}
\label{ea01}%
\end{equation}
From Eqs. (\ref{e01}) and (\ref{ea01}), we obtain a useful relation between
$X$ and $Y$:%
\begin{equation}
X\sim\left(  \varepsilon e_{0}^{-\left(  1/n_{h}-1/n_{s}\right)  }\right)
^{\frac{1}{1+1/n_{s}}}Y\equiv\tilde{\varepsilon}Y
\end{equation}

This implies that $(X,Y)=(\tilde{\varepsilon}L,L)$ when $a/\tilde{\varepsilon
}>L$ and $(X,Y)=(a,a/\tilde{\varepsilon})$ otherwise. This is because the
natural characteristic length $X$ is $a$ if $L>>a$, but, if $L$ is not so
larger than $a$ and if the length scale $Y\sim a/\tilde{\varepsilon}$ is
cutoff by $L$ (i.e., if $a/\tilde{\varepsilon} \gtrsim L$), then $Y$ should
scale as $L$, which means $X\sim\tilde{\varepsilon}L$. In the case of the
monolithic hard material, $X_{h}\sim Y_{h}$ so that $(X_{h},Y_{h})=(L,L)$ when
$a\gtrsim L$ and $(X_{h},Y_{h})=(a,a)$ otherwise.

The enhancement factor for the fracture stress is obtained as%
\begin{equation}
\frac{\sigma_{F}}{\sigma_{F}^{MN}}=\left(  \frac{d}{a_{0}}\varepsilon
^{-\frac{1}{1+1/n_{s}}}\left(  \frac{G_{h}/E_{h}}{a_{0}}\right)  ^{\frac
{1}{1+n_{h}}+\frac{1}{1+1/n_{s}}-1}\frac{X_{h}}{Y}\right)  ^{\frac{1}{1+n_{h}%
}} \label{e07b}%
\end{equation}
This can be obtained in the following three ways, which are physically the
same: (I) Substitute Eq.~(\ref{e07}) into Eq.~(\ref{e05}). (II) Match
Eq.~(\ref{ea2}) evaluated at $r=d$ with $\sigma_{h}$ and interpret the remote
stress $\sigma_{0}$ in Eq.~(\ref{ea2}) as the fracture strength $\sigma_{F}$.
(III) Calculate $\sigma_{0}u_{0}$ where $u_{0}$ is given by $u$ in Eq.
(\ref{ea01}) and identify $\sigma_{0}$ as $\sigma_{F}$.

When the crack is large so that $X_{h}=Y=L$, we obtain Eq.~(\ref{em2}) from
Eq.~(\ref{e07b}). Equation (\ref{em5}) is obtained for such large cracks by
setting $r=d$ and $X=\tilde{\varepsilon}L$ in Eq. (\ref{ea2}), with using Eq.
(\ref{ea01}). While Eqs. (\ref{em1}) and (\ref{em2}) are valid only at the
critical state of failure, Eq. (\ref{em5}) is valid as long as $\sigma_{0}$
does not exceed the critical stress $\sigma_{F}$.

When the crack is small so that $X_{h}=a$ and $Y=a/\tilde{\varepsilon}$, we
obtain from Eq.~(\ref{e07b}):%
\begin{equation}
\frac{\sigma_{F}}{\sigma_{F}^{MN}}=\left(  \frac{d}{a_{0}}\left(  \frac
{G_{h}/E_{h}}{a_{0}}\right)  ^{\frac{1}{1+n_{h}}+\frac{1}{1+1/n_{s}}-1}\left(
e_{0}^{-\left(  1/n_{h}-1/n_{s}\right)  }\right)  ^{\frac{1}{1+1/n_{s}}%
}\right)  ^{\frac{1}{1+n_{h}}}%
\end{equation}
For such small cracks, by setting $r=d$ and $X=a$ in Eq. (\ref{ea2}), with
using Eq. (\ref{ea01}).
\begin{equation}
\sigma_{M}/E_{h}\sim e_{0}^{1/n_{h}}(a/d)^{\eta}%
\end{equation}

The large crack assumption may be satisfied in our numerical calculations.
Note that for $\varepsilon\simeq0.001$ and $e_{0}=0.008$, then $\tilde
{\varepsilon}\simeq0.1$ so that $a>k\tilde{\varepsilon}L$ holds if $k\,\ $is
slightly smaller $1/3$ (i.e., $k$ is of the order of unity) even if
$L/d_{0}=10000$, which is the largest value we used in numerical calculation
because $a/d_{0}=320$. (If otherwise, $\sigma_{M}$ would be $L/d$-independent.)


\end{document}